\documentclass[pra,twocolumn,showpacs]{revtex4}
\usepackage{graphicx}
\usepackage{dcolumn}
\usepackage{amsmath}
\usepackage{amssymb}
\newcommand{\bfr}{\mathbf{r}}
\newcommand{\bfu}{\mathbf{u}}

\begin{document}

\title{Exact-exchange energy density in the gauge of a semilocal
density functional approximation}
\author{Jianmin Tao}
\affiliation{Theoretical Division, Los Alamos National Laboratory,
Los Alamos, New Mexico 87545, USA}
\author{Viktor N. Staroverov}
\affiliation{Department of Chemistry, University of Western Ontario,
London, Ontario N6A 5B7, Canada}
\author{Gustavo E. Scuseria}
\affiliation{Department of Chemistry, Rice University, Houston,
Texas 77005, USA}
\author{John P. Perdew}
\affiliation{Department of Physics and Quantum Theory Group,
Tulane University, New Orleans, Louisiana 70118, USA}

\date{\today}

\begin{abstract}
Exact-exchange energy density and energy density of a semilocal
density functional approximation are two key ingredients for modeling
the static correlation, a strongly nonlocal functional of the electron
density, through a local hybrid functional.  Because energy densities
are not uniquely defined, the conventional (Slater) exact-exchange
energy density $e_\mathrm{x}^\mathrm{ex(conv)}$ is not necessarily
well-suited for local mixing with a given semilocal approximation.
We show how to transform $e_\mathrm{x}^\mathrm{ex(conv)}$ in order to
make it compatible with an arbitrary semilocal density functional,
taking the nonempirical meta-generalized gradient approximation
of Tao, Perdew, Staroverov, and Scuseria (TPSS) as an example.
Our additive gauge transformation function integrates to zero,
satisfies exact constraints, and is most important where the
density is dominated by a single orbital shape.
We show that, as expected, the difference between semilocal and
exact-exchange energy densities becomes more negative under bond
stretching in He$_2^{+}$ and related systems. Our construction
of $e_\mathrm{x}^\mathrm{ex(conv)}$ by a resolution-of-the-identity
method requires uncontracted basis functions.
\end{abstract}

\maketitle

\section{Introduction}

In Kohn-Sham density functional
theory~\cite{Kohn:1965/PR/A1133,Book/Fiolhais:2003},
the exchange-correlation (xc) energy $E_\mathrm{xc}$ must be approximated
as a functional of the electron spin-densities $n_\uparrow(\bfr)$
and $n_\downarrow(\bfr)$. This functional can be always written as
\begin{equation} 
 E_\mathrm{xc}[n_\uparrow,n_\downarrow]
  = \int d\bfr\, e_\mathrm{xc}(\bfr),  \label{eq:exc}
\end{equation}
where $e_\mathrm{xc}(\bfr)=n(\bfr)\varepsilon_\mathrm{xc}(\bfr)$ is
the exchange-correlation energy density, $n=n_\uparrow+n_\downarrow$
is the total electron density, and $\varepsilon_\mathrm{xc}$ is
the exchange-correlation energy per electron.  Approximations to
$\varepsilon_\mathrm{xc}(\bfr)$ can be constructed in a fairly
systematic way~\cite{Perdew:2001/1,Perdew:2005/JCP/062201} by
employing increasingly complex ingredients built from the Kohn-Sham
orbitals. Most of the existing exchange-correlation approximations
use only ingredients found from the occupied Kohn-Sham orbitals
at $\bfr$ or in an infinitesimal neighborhood of $\bfr$, such as
$n_\sigma(\bfr)=\sum_{i}^\mathrm{occ.}|\phi_{i\sigma}(\bfr)|^2$,
$\nabla n_\sigma(\bfr)$, and
$\tau_\sigma=\frac{1}{2}\sum_{i}^\mathrm{occ.}
|\nabla\phi_{i\sigma}(\bfr)|^2$, where
$\sigma=\uparrow,\downarrow$.  Such functionals are
called semilocal and include the local spin density
approximation~\cite{Kohn:1965/PR/A1133,Vosko:1980/CJP/1200,%
Perdew:1992/PRB/13244}, the generalized gradient approximation
(GGA)~\cite{Perdew:1996/PRL/3865},
and the meta-GGA~\cite{Tao:2003/PRL/146401}.

Semilocal functionals are often
accurate~\cite{Staroverov:2003/JCP/12129,%
Staroverov:2004/PRB/075102,Csonka:2005/IJQC/506,Tao:2007/PM/1071,%
Furche:2006/JCP/044103} but tend to make large
errors for open systems of fluctuating electron
number~\cite{Perdew:1985/265,Perdew:1990/AQC/113,%
Mori-Sanchez:2006/JCP/201102,Ruzsinszky:2007/JCP/104102,Vydrov:2007/JCP/154109},
such as fragments connected by stretched bonds.  This occurs
because semilocal functionals respect the exact hole sum rule for
a closed system but not for an open one of fluctuating electron
number~\cite{Perdew:2007/PRA/040501}, where (after symmetry breaking)
the semilocal exchange typically overestimates the magnitude of the
static correlation.  We have argued~\cite{Perdew:2007/PRA/040501}
that, in order to correct these errors, one needs to go
beyond the semilocal approximation and incorporate a fully
nonlocal ingredient, the exact-exchange (ex) energy density
$e_\mathrm{x}^\mathrm{ex}(\bfr)$ conventionally (conv) defined as
\begin{equation}
 e_\mathrm{x}^\mathrm{ex(conv)}(\bfr) =
 -\frac{1}{2} \sum_{\sigma=\uparrow,\downarrow} \int d\bfr' \,
 \frac{|\gamma_\sigma(\bfr,\bfr')|^2}{|\bfr-\bfr'|},  \label{eq:ex-conv}
\end{equation}
where $\gamma_\sigma(\bfr,\bfr')$ is the one-electron $\sigma$-spin
density matrix of the Kohn-Sham reference system
\begin{equation}
 \gamma_\sigma(\bfr,\bfr') = 
 \sum_i^\mathrm{occ.} \phi_{i\sigma}(\bfr)\phi_{i\sigma}^*(\bfr'). 
    \label{eq:dm1}
\end{equation}
In the Jacob's ladder classification of density functional
approximations~\cite{Perdew:2001/1},
functionals that employ $e_\mathrm{x}^\mathrm{ex}(\bfr)$
are called hyper-GGAs. We have also argued that a hyper-GGA
can simultaneously achieve good accuracy and satisfy
important exact constraints if 
the exact-exchange energy density is combined
with a semilocal (sl) exchange-correlation in
a so-called local hybrid (lh) functional
\begin{equation}
 e_\mathrm{xc}^\mathrm{lh}
  = e_\mathrm{x}^\mathrm{ex} +
 [1 - a(\bfr)](e_\mathrm{x}^\mathrm{sl}
 - e_\mathrm{x}^\mathrm{ex}) + e_\mathrm{c}^\mathrm{sl}, \label{eq:lh}
\end{equation}
where $0\le a(\bfr)\le 1$ is the position-dependent mixing
function. If $a(\bfr)=\mathrm{const}$, Eq.~(\ref{eq:lh}) reduces to
a global hybrid (gh) functional~\cite{Becke:1993/JCP/5648}.
The general local hybrid form was suggested
by Cruz \textit{et al.}~\cite{Cruz:1998/JPCA/4911}
as early as 1998, but specific forms of $a(\bfr)$
were not proposed until later~\cite{Perdew:2001/1,%
Jaramillo:2003/JCP/1068}. The fundamental physical justification
for local hybrids has been advanced only recently~\cite{Perdew:2007/PRA/040501}.
The local hybrid approach is, of course, not the only way of
attacking the static correlation problem. Other distinct approaches
are being actively pursued~\cite{Becke:2005/JCP/064101,%
Mori-Sanchez:2006/JCP/091102}.

When $a(\bfr)$ of Eq.~(\ref{eq:lh}) tends to 1 in the high-density
limit, the local hybrid functional uses full exact
exchange and treats correlation as the sum of two parts, the static
(long-range, left-right) and dynamic (short-range) correlation.
The dynamic correlation is relatively easy to model 
by a semilocal correlation functional
$e_\mathrm{c}^\mathrm{sl}(\bfr)$. The static correlation
is represented by the difference
$[e_\mathrm{x}^\mathrm{sl}(\bfr)-e_\mathrm{x}^\mathrm{ex}(\bfr)]$
weighted by a position-dependent function $[1-a(\bfr)]$.  This
form is motivated by evidence that some (typically more than 100\%)
of the static correlation is already contained in semilocal exchange
approximations~\cite{Perdew:2007/PRA/040501,Schipper:1998/PRA/1729,%
Handy:2001/MP/403,Molawi:2002/IJQC/86},
but not in $e_\mathrm{x}^\mathrm{ex}(\bfr)$.

Any proposal for a practical local hybrid functional must
deal with the fact that, while the total energy is measurable,
physical, and unique, the energy \textit{density} is not.  As a
result, an arbitrary function $G(\bfr)$ that has a dimension of
energy per volume and integrates to zero can be added to any
energy density of a global hybrid functional with no effect
on the total energy.  In contrast, addition of $G(\bfr)$ to
$e_\mathrm{x}^\mathrm{ex}(\bfr)$ or $e_\mathrm{x}^\mathrm{sl}(\bfr)$
in a local hybrid of Eq.~(\ref{eq:lh}) \textit{will} affect the
total energy because the exact-exchange energy density here is
weighted locally.

While there is no ``most correct'' choice for the xc-energy density,
there is indeed a conventional choice, which for
exchange is Eq.~(\ref{eq:ex-conv}). However, the conventional exact
exchange energy density does not have a second-order gradient
expansion~\cite{Armiento:2002/PRB/165117} and so is not the most
natural choice for density-functional approximation.  Also, the
exchange hole associated with $e_\mathrm{x}^\mathrm{ex(conv)}$
is highly delocalized, which makes it very difficult to model with
semilocal functionals.  Standard functionals are at most designed
to recover the conventional (or any other) exchange energy density
to zeroth-order in the density gradients, i.e., for uniform electron
densities only.

In the local hybrids proposed to date~\cite{Jaramillo:2003/JCP/1068,%
Arbuznikov:2006/JCP/204102,Bahmann:2007/JCP/011103,Arbuznikov:2007/CPL/160,%
Janesko:2007/JCP/164117}, $e_\mathrm{x}^\mathrm{sl}(\bfr)$ is taken as the
integrand of the semilocal functional $E_\mathrm{x}^\mathrm{sl}$
as written, while $e_\mathrm{x}^\mathrm{ex}$ is taken
as $e_\mathrm{x}^\mathrm{ex(conv)}$. This choice
is not necessarily the one best suited
for modeling the static correlation by the difference
$(e_\mathrm{x}^\mathrm{sl}-e_\mathrm{x}^\mathrm{ex})$. Moreover,
the very idea of attaching physical significance to the difference
$(e_\mathrm{x}^\mathrm{sl}-e_\mathrm{x}^\mathrm{ex})$ requires that
both $e_\mathrm{x}^\mathrm{sl}$ and $e_\mathrm{x}^\mathrm{ex}$
be defined with respect to some common
reference or gauge.

The choice of the gauge itself is a matter of convention.
One such choice is based on the
Levy-Perdew virial relation~\cite{Levy:1985/PRA/2010}.
Burke~\textit{et al.}~\cite{Burke:1998/JCP/8161} have
pointed out that virial exchange energy densities
$e_\mathrm{x}^\mathrm{vir}(\bfr)=-n(\bfr)\bfr\cdot\nabla v_\mathrm{x}(\bfr)$,
where $v_\mathrm{x}(\bfr)=\delta E_\mathrm{x}/\delta n(\bfr)$,
are unique for any given functional. However, the virial
energy density depends on the choice of
origin of $\bfr$ and has other undesirable properties.
Furthermore, for the exact-exchange energy density,
this approach requires constructing the
optimized effective potential~\cite{Talman:1976/PRA/36,%
Kummel:2003/PRL/043004} (OEP),
a procedure that is problematic in finite basis
sets~\cite{Staroverov:2006/JCP/141103,Staroverov:2006/JCP/081104,
Izmaylov:2007/JCP/084113}.
Burke~\textit{et al.}~have also proposed~\cite{Burke:1998/JCP/8161}
and investigated~\cite{Burke:1998/IJQC/583} the ``unambiguous"
exchange-(correlation) energy density which
is uniquely determined by the corresponding
energy functional via the exchange-(correlation) potential
and the Helmholtz theorem.
This ``unambiguous" exact-exchange energy density has all the
desired properties but, like the virial energy density,
requires construction of the OEP and, hence, is not
very practical at present.

In this work, we propose  and implement two new, dependable
methods in which $e_\mathrm{x}^\mathrm{sl}(\bfr)$ serves as the
reference and $e_\mathrm{x}^\mathrm{ex}(\bfr)$ is ``tuned" to
the gauge of $e_\mathrm{x}^\mathrm{sl}(\bfr)$. The first of them,
summarized in section~\ref{sec:vecf} below, is the one we will use
in a still-unpublished hyper-GGA~\cite{HGGA:xxx} based upon the
ideas of Ref.~\cite{Perdew:2007/PRA/040501}.

\section{Theory}

For a slowly-varying electron density, the conventional
exact-exchange energy density will be well approximated by
local or semilocal density functionals, although (unlike
the integrated exchange energy) it has no analytic gradient
expansion~\cite{Perdew:1988/MAS,Armiento:2002/PRB/165117}.
Our idea for making $e_\mathrm{x}^\mathrm{ex}$ compatible with
a given $e_\mathrm{x}^\mathrm{sl}$ is based on the observation
that although the static correlation is generally quite large
(comparable in magnitude to exchange), it is negligible
in compact closed systems, such as atoms with nondegenerate
electron configurations. Therefore, $e_\mathrm{x}^\mathrm{ex}$
should be close to $e_\mathrm{x}^\mathrm{sl}$ at each $\bfr$
in such systems. This is consistent with the fact that the
conventional exact-exchange~\cite{Folland:1971/PRA/1535,%
Becke:1988/PRA/3098,Tao:2001/JCP/3519} or
exchange-correlation~\cite{Cancio:2006/PRB/081202} energy densities
in compact closed systems can often be modeled very accurately
using only semilocal ingredients.

Hence, we will make $e_\mathrm{x}^\mathrm{ex}$ as close as
possible to $e_\mathrm{x}^\mathrm{sl}$ in those systems where the static
correlation is known to be small. This can be achieved by various
means: for example, by adding to $e_\mathrm{x}^\mathrm{ex(conv)}$
a term that integrates to zero. We say that
the resulting exact-exchange energy density is in the \textit{gauge}
of that particular semilocal exchange approximation and denote it
by $e_\mathrm{x}^\mathrm{ex(sl)}$.  To illustrate this method,
we will construct the exact-exchange energy density in the
gauge of the meta-GGA of Tao, Perdew, Staroverov, and Scuseria
(TPSS)~\cite{Tao:2003/PRL/146401}.

\subsection{Construction from the divergence of a vector field}
\label{sec:vecf}

For use as $e_\mathrm{x}^\mathrm{ex}(\bfr)$ in Eq.~(\ref{eq:lh}),
we construct the exact-exchange energy density in the gauge of
a semilocal functional as follows. First we write
\begin{equation}
 e_\mathrm{x}^\mathrm{ex(sl)}(\bfr) = 
 e_\mathrm{x}^\mathrm{ex(conv)}(\bfr) + G(\bfr),  \label{eq:ex-tpss}
\end{equation}
where $e_\mathrm{x}^\mathrm{ex(conv)}(\bfr)$ is the conventional
exact-exchange energy density given by Eq.~(\ref{eq:ex-conv})
and $G(\bfr)$ is the gauge transformation term to be determined,
such that
\begin{equation}
 \int d\bfr\,G(\bfr) = 0.  \label{eq:per1}
\end{equation}
Obviously, Eq.~(\ref{eq:per1}) leaves much freedom in choosing the
analytic form of function $G(\bfr)$. The range of possibilities
can be narrowed down by several physical considerations:
a) $e_\mathrm{x}^\mathrm{ex(sl)}(\bfr)$ should reproduce
$e_\mathrm{x}^\mathrm{sl}(\bfr)$ in atoms as closely as possible;
b) for use in a hyper-GGA, $G(\bfr)$ should contain only
the hyper-GGA ingredients, i.e., $n_\sigma(\bfr)$,
$\tau_\sigma=\frac{1}{2}\sum_{i}^\mathrm{occ.}
|\nabla\phi_{i\sigma}(\bfr)|^2$,
$e_\mathrm{x\sigma}^\mathrm{ex(conv)}(\bfr)$, and, possibly,
their derivatives; c) $e_\mathrm{x}^\mathrm{ex(sl)}(\bfr)$ should
satisfy as many exact constraints as possible.

We start the construction of $G(\bfr)$ for spin-unpolarized
systems by noting that the integral of the divergence of any
well-behaved rapidly decaying vector field $\mathbf{F}(\bfr)$
is zero, that is, $\int d\bfr\,\nabla\cdot\mathbf{F}(\bfr)=0$.
So we will take $G(\bfr)=\nabla\cdot\mathbf{F}(\bfr)$. The vector
field $\mathbf{F}(\bfr)$ itself will be chosen from the requirement
that $e_\mathrm{x}^\mathrm{ex(conv)}(\bfr)+G(\bfr)$ satisfy the most
basic properties of the exchange energy density: correct coordinate
scaling, finiteness at the nucleus, etc.

One particular form that meets these requirements is:
\begin{equation}
 G(\bfr)= a \nabla \cdot \left[
 \frac{n/\tilde{\varepsilon}^{2}}{1+c\left(n/\tilde{\varepsilon}^3\right)^2}
 \left(\frac{\tau^W}{\tau}\right)^b \nabla\tilde{\varepsilon} \right],
  \label{eq:delta}
\end{equation}
where $\tilde{\varepsilon}(\bfr)
=-\varepsilon_\mathrm{x}^\mathrm{ex(conv)}(\bfr)$,
$\tau^W=|\nabla n|^2/8n$ is the von
Weizs\"acker~\cite{Weizsacker:1935/ZP/431} kinetic energy density for
real orbitals, $\tau=\tau_\uparrow+\tau_\downarrow$ is the Kohn-Sham
kinetic energy density, and $a$, $b$, and $c$ $(c>0)$ are adjustable
parameters. Note that $0\leq \tau^W/\tau \leq 1$~\cite{Kurth:1999/IJQC/889}.

The function $G(\bfr)$ of Eq.~(\ref{eq:delta}) has the following
exact properties of the exact-exchange energy density in
the conventional gauge (or coordinate-transformed as described
in Sec.~\ref{sec:t-hole}):

(i) Correct uniform coordinate scaling. Under this transformation,
the conventional exact-exchange energy density
behaves~\cite{Levy:1985/PRA/2010} like
$e_\mathrm{x}^\mathrm{ex(conv)}(\bfr)
=\lambda^4 e_\mathrm{x}^\mathrm{ex(conv)}(\lambda\bfr)$
or, in shorthand, $e_\mathrm{x}^\mathrm{ex(conv)} \sim\lambda^4$.
The ingredients of $G(\bfr)$ behave like
$n\sim\lambda^3$, $\tilde{\varepsilon}\sim\lambda$, $\tau^W\sim\lambda^5$, 
$\tau\sim\lambda^5$, $\nabla\sim\lambda$, so
$G_\lambda(\bfr) = \lambda^4 G(\lambda \bfr)$,
which is the correct behavior.

(ii) Correct nonuniform coordinate scaling~\cite{Levy:1991/PRA/4637}.
Under this scaling, the density behaves like
$n_\lambda^x (\bfr) = \lambda n(\lambda x,y,z)$ or, in shorthand,
$n\sim\lambda$. The other ingredients scale in the $\lambda\to\infty$ limit
like $\tilde{\varepsilon}\sim\lambda^0$, $\tau^W\sim\lambda^3$, $\tau\sim\lambda^3$,
$\nabla\sim\lambda$, so in this limit
$G_\lambda^x(x,y,z) = \lambda G(\lambda x,y,z)$,
which is the correct nonuniform coordinate scaling property of the exchange
energy density.

(iii) $G(\bfr)$ is finite everywhere. This is because
$\tilde{\varepsilon}$ has no cusp at the nucleus~\cite{March:2000/PRA/012520},
which ensures that $\nabla^2\tilde{\varepsilon}$ is finite. All other ingredients
of $G$ are also finite.

(iv) $G(\bfr)$ vanishes for a uniform electron gas and, more
generally, satisfies Eq.~(\ref{eq:per1}).

We also note that, at large $r$, the density decays exponentially,
$n\sim e^{-\alpha r}$, where $\alpha$ is a constant,
$\tau^W/\tau\to 1$, $\tilde{\varepsilon}\sim 1/r$,
so the large-$r$ behavior is $G(r)\sim -\partial n/\partial r \sim n$,
which is comparable to the $-n/2r$ decay of $e_\mathrm{x}^\mathrm{ex(conv)}$.

The values of $a$, $b$, and $c$ are determined by fitting
$e_\mathrm{x}^\mathrm{ex(conv)}(\bfr)+G(\bfr)$ to
$e_\mathrm{x}^\mathrm{sl}(\bfr)$, where sl=TPSS. In doing so, we
note that for one- and closed-shell two-electron (iso-orbital)
densities $\tau^W/\tau=1$, so $G(\bfr)$ is fixed by the
parameters $a$ and $c$ alone. We use two model-atom iso-orbital
densities: the exact two-electron exponential density $n(r) = (2/\pi)e^{-2r}$
and the two-electron cuspless density $n(r) =
(1/2\pi)(1 + 2r)e^{-2r}$. In the case of sl=TPSS, the fit gives
$a=0.015$ and $c=0.04$. The value $b=4$ is chosen to
be an integer that gives the best fit to the TPSS exchange energy
density for the 8-electron jellium cluster with $r_s=4$ bohr.
This choice ensures that $G(\bfr)$ is very small (of the order
of $\nabla^{10}$) for a slowly varying density, as it should be.
Gauge corrections for semilocal functionals other than TPSS can be
constructed similarly by assuming the same analytic form for $G(\bfr)$
and refitting the parameters $a$, $b$, and $c$.

While we cannot rule out that there exists a simpler function
$G(\bfr)$ that satisfies exact constraints (i)--(iv), we can
point out that many obvious candidates definitely fail to do so.
For example, the function $\nabla^2 n^{2/3}$, motivated by the
work of Cancio and Chou~\cite{Cancio:2006/PRB/081202}, correctly
integrates to zero and has the correct uniform scaling property,
but diverges at the nucleus and does not have the proper nonuniform
scaling property.

For a partly or fully spin-polarized system, the gauge correction
becomes the sum of same-spin contributions
$G(\bfr)=\sum_\sigma G_\sigma(\bfr)$.
To deduce the form of $G_\sigma(\bfr)$ we use
the spin scaling relation~\cite{Oliver:1979/PRA/397}:
\begin{equation}
 E_\mathrm{x}[n_\uparrow,n_\downarrow] =
 \frac{1}{2}E_\mathrm{x}[2n_\uparrow]
  + \frac{1}{2}E_\mathrm{x}[2n_\downarrow],  \label{eq:ss}
\end{equation}
which also holds for exchange energy densities.
Applying Eq.~(\ref{eq:ss}) to $G(\bfr)$, we write
\begin{equation}
 G([n_\uparrow,n_\downarrow];\bfr) = 
 \frac{1}{2}\sum_\sigma G([2 n_\sigma];\bfr)
\end{equation}
and define $G_\sigma(\bfr) \equiv \frac{1}{2}
G([2 n_\sigma];\bfr)$. Thus, for spin-polarized systems
$G(\bfr)=\sum_\sigma G_\sigma(\bfr)$, where
\begin{equation}
 G_\sigma(\bfr) = a \nabla \cdot \left[
  \frac{n_\sigma/\tilde{\varepsilon}_\sigma^{2}}
 {1 + 4c\left(n_\sigma/\tilde{\varepsilon}_\sigma^3\right)^2}
 \left(\frac{\tau_\sigma^W}{\tau_\sigma}\right)^b
  \nabla \tilde{\varepsilon}_\sigma \right],
  \label{eq:spindelta}
\end{equation}
in which $\tau_\sigma^W=|\nabla n_\sigma|^2/8n_\sigma$ and
\begin{equation}
 \tilde{\varepsilon}_\sigma=-\varepsilon_\mathrm{x\sigma}^\mathrm{ex(conv)}
 = -\frac{e_\mathrm{x\sigma}^\mathrm{ex(conv)}}{n_\sigma}.
   \label{eq:vareps}
\end{equation}
Note that $\varepsilon_\mathrm{x}^\mathrm{ex}\neq
\sum_\sigma\varepsilon_\mathrm{x\sigma}^\mathrm{ex}$
but $e_\mathrm{x}^\mathrm{ex}=\sum_\sigma e_\mathrm{x\sigma}^\mathrm{ex}$
because the spin-scaling relation~(\ref{eq:ss}) applies
only to energy densities.

\subsection{Construction by a coordinate transformation
of the exact-exchange hole}
\label{sec:t-hole}

The exact-exchange energy density can be also converted to the gauge of
a semilocal approximation by transforming the exact-exchange hole.
The conventional exact-exchange energy density can be written as
\begin{equation}
 e_\mathrm{x\sigma}^\mathrm{ex(conv)}(\bfr)
  = \frac{n_\sigma(\bfr)}{2} \int d\bfr' \,
 \frac{h_\mathrm{x\sigma}(\bfr,\bfr')}{|\bfr-\bfr'|}, \label{eq:spin}
\end{equation}
where $h_\mathrm{x\sigma}(\bfr,\bfr')$ is the exact-exchange hole 
\begin{equation}
 h_\mathrm{x\sigma}(\bfr,\bfr')
  = -\frac{|\gamma_{\sigma}(\bfr,\bfr')|^2}{n_\sigma(\bfr)},
  \label{eq:x-hole}
\end{equation}
This hole is highly delocalized but can be made less
so~\cite{Koehl:1996/MP/835,Springborg:1999/CPL/83,Tao:2003/JCP/6457} by
an appropriate coordinate transformation
$(\bfr,\bfr')\to(\bfr_1,\bfr_2)$ of the density
matrix, such as~\cite{Tao:2003/JCP/6457}
\begin{equation}
 \left( \begin{array}{cc}
 \bfr \\ \bfr' \end{array} \right)
  = \left( \begin{array}{cc}
 2 - \omega & -1 + \omega \\
 1 - \omega & \omega \end{array} \right)
 \left( \begin{array}{cc}
 \bfr_1 \\ \bfr_2 \end{array} \right),   \label{eq:coord}
\end{equation}
where $0<\omega<1$.
This transformation does not affect the total exchange energy
$E_\mathrm{x}$ but yields a distinctly different exchange energy density
\begin{equation}
 e_\mathrm{x\sigma}^\mathrm{ex(\omega)}(\bfr_1)
  = \frac{n_\sigma(\bfr_1)}{2} \int d\bfu \,
 \frac{h_\mathrm{x\sigma}^\omega(\bfr_1,\bfr_1+\bfu)}{u},
  \label{eq:ex-t}
\end{equation}
where $\bfu=\bfr_2-\bfr_1$ and
$h_\mathrm{x\sigma}^\omega(\bfr_1,\bfr_1+\bfu)$
is the transformed exact-exchange hole~\cite{Tao:2003/JCP/6457}
\begin{eqnarray}
 h_\mathrm{x\sigma}^\omega(\bfr_1,\bfr_1+\bfu)
 & = & h_\mathrm{x\sigma}(\bfr_1+[\omega-1]\bfu,\bfr_1+\omega\bfu)
 \nonumber \\
 & & \times \frac{n_\sigma(\bfr_1+[\omega-1]\bfu)}{n_\sigma(\bfr_1)},
  \label{eq:thole}
\end{eqnarray}
given in terms of the conventional exchange hole.  Since the
exchange hole associated with a semilocal functional is relatively
local, a transformation of the exact-exchange hole by
Eq.~(\ref{eq:thole}) can make $e_\mathrm{x\sigma}^\mathrm{ex(\omega)}$
resemble its semilocal approximation more closely than
$e_\mathrm{x\sigma}^\mathrm{ex(conv)}$ does.
It should be noted that the transformed
hole of Eq.~(\ref{eq:ex-t}) does not obey the sum rule for the
conventional hole at each $\bfr_1$, but preserves the correct
normalization of the
system-averaged exchange hole~\cite{Tao:2003/JCP/6457}.

The extent of locality of the exchange hole depends on the value
of parameter $\omega$. The maximal localization is achieved
at $\omega=1/2$~\cite{Tao:2003/JCP/6457,Springborg:2001/ZPC/1243}.  We have
numerically evaluated the transformed exact-exchange energy density
$e_\mathrm{x\sigma}^\mathrm{ex(\omega)}(\bfr)$ for various values
of $\omega$ and found that $\omega=0.92$ leads to the best fit of the
exact-exchange energy density to the TPSS meta-GGA.

\section{Computational methodology}

In practice, it is much easier to construct the gauge correction\
function $G(\bfr)$ than to perform numerical integration over
transformed coordinates of the exchange hole in Eq.~(\ref{eq:ex-t}).
Therefore, we will adopt the former method for
the purpose of constructing a hyper-GGA functional. In this
section, we describe a general-purpose implementation of the
TPSS gauge term $G(\bfr)$ in finite basis sets.

\subsection{Evaluation of the exact-exchange energy density
in the conventional gauge}

Analytic evaluation of the conventional exact-exchange energy
density by Eqs.~(\ref{eq:ex-conv}) and (\ref{eq:dm1}) is possible but
impractical because it requires evaluation and contraction of many 
one-electron integrals for each grid point $\bfr$. Instead, we employ
a much more efficient approximate method of
Della Sala and G\"{o}rling~\cite{DellaSala:2001/JCP/5718}.
Although this method is documented in the
literature~\cite{DellaSala:2001/JCP/5718,Arbuznikov:2006/JCP/204102},
we will supply its detailed derivation here because it serves as
a stepping-stone for evaluating our function $G(\bfr)$.

When a basis set $\{\chi_\mu\}$ is introduced, each Kohn-Sham orbital
is taken as a linear combination of one-electron basis functions,
$\phi_{i\sigma}(\bfr)=\sum_\mu c_{\mu i}^\sigma \chi_\mu(\bfr)$.
In terms of these basis functions, the density matrix of
Eq.~(\ref{eq:dm1}) is
\begin{equation}
 \gamma_\sigma(\bfr,\bfr')
 = \sum_{\mu\nu} P_{\mu\nu}^\sigma \chi_\mu(\bfr) \chi_\nu^*(\bfr'),
\end{equation}
where $P_{\mu\nu}^\sigma=P_{\nu\mu}^\sigma=\sum_i^\mathrm{occ.}
c_{\mu i}^\sigma (c_{\nu i}^\sigma)^*$. The conventional
exact-exchange energy density of Eq.~(\ref{eq:ex-conv})
can be written as
\begin{eqnarray}
 e_\mathrm{x\sigma}^\mathrm{ex(conv)}(\bfr)
 & = & -\frac{1}{2} \sum_{\eta\kappa} \sum_{\rho\nu} \int d\bfr'\,
 P_{\eta\kappa}^\sigma P_{\rho\nu}^\sigma \nonumber \\
 & & \times \frac{\chi_\eta(\bfr) \chi_\nu^*(\bfr)
  \chi_\rho(\bfr') \chi_\kappa^*(\bfr')}{|\bfr-\bfr'|}.
 \label{eq:ex-P}
\end{eqnarray}
The single integral over $\bfr'$ in Eq.~(\ref{eq:ex-P}) is not
so easily evaluated for many different values of $\bfr$,
but introducing a second integration over $\bfr$ yields
$E_\mathrm{x\sigma}^\mathrm{ex}$, which is evaluated analytically
and simply in Gaussian basis sets. These facts motivate the following
development. Using the $\delta$-function one can write
\begin{equation}
 \frac{\chi_\eta(\bfr)}{|\bfr-\bfr'|}
 = \int d\bfr''\, \frac{\chi_\eta(\bfr'')}{|\bfr''-\bfr'|}
 \delta(\bfr''-\bfr).  \label{eq:delta-act}
\end{equation}
The $\delta$-function can be approximated by an expansion
in the same non-orthogonal basis as the orbitals, namely,
$\delta(\bfr''-\bfr)=\sum_\mu \chi_\mu(\bfr)c_\mu(\bfr'')$,
whose Fourier coefficients $c_\mu(\bfr')$ can be determined as usual.
This yields
\begin{equation}
 \delta(\bfr''-\bfr) = \sum_\mathrm{\mu\xi}
 \chi_\mu(\bfr) S_{\mu\xi}^{-1} \chi_\xi^*(\bfr''),
  \label{eq:delta-f}
\end{equation}
where $S_{\mu\xi}^{-1}$ are matrix elements of the inverse
of the basis set overlap matrix.
Substitution of Eqs.~(\ref{eq:delta-act}) and (\ref{eq:delta-f})
into Eq.~(\ref{eq:ex-P}) gives
\begin{equation}
 e_\mathrm{x\sigma}^\mathrm{ex(conv)}(\bfr)
 = \frac{1}{2} \sum_{\rho\nu} \sum_{\mu\xi}
 S_{\mu\xi}^{-1} K_{\xi\rho}^\sigma 
  P_{\rho\nu}^\sigma \chi_\mu(\bfr)\chi_\nu^*(\bfr),
 \label{eq:ex-RI1}
\end{equation}
where
\begin{equation}
 K_{\xi\rho}^\sigma = 
 -\sum_{\eta\kappa} P_{\eta\kappa}^\sigma
 \int d\bfr' \int d\bfr''\,
  \frac{\chi_\rho(\bfr')\chi_\kappa^*(\bfr')
 \chi_\eta(\bfr'')\chi_\xi^*(\bfr'')}{|\bfr''-\bfr'|}
\end{equation}
are elements of the exchange matrix. Eq.~(\ref{eq:ex-RI1})
can be rewritten as
\begin{equation}
 e_\mathrm{x\sigma}^\mathrm{ex(conv)}(\bfr)
  = \frac{1}{2} \sum_{\mu\nu} \tilde{Q}_{\mu\nu}^\sigma
 \chi_\mu(\bfr)\chi_\nu^*(\bfr),  \label{eq:ex-RI2}
\end{equation}
where $\tilde{Q}_{\mu\nu}^\sigma$ are elements of the matrix
$\tilde{\mathbf{Q}}^\sigma=\mathbf{S}^{-1}\mathbf{K}^\sigma\mathbf{P}^\sigma$.
Eq.~(\ref{eq:ex-RI2}) is analogous to the formula for the density
$n(\bfr)\equiv\gamma_\sigma(\bfr,\bfr)=\sum_{\mu\nu} P_{\mu\nu}^\sigma
 \chi_\mu(\bfr)\chi_\nu^*(\bfr)$ except that, unlike $\mathbf{P}^\sigma$,
the matrix $\tilde{\mathbf{Q}}^\sigma$ is generally not symmetric.
The analogy can be made complete by replacing $\tilde{\mathbf{Q}}^\sigma$
with the symmetrized matrix
\begin{equation}
 \mathbf{Q}^\sigma = \frac{1}{2} \left(
  \mathbf{P}^\sigma\mathbf{K}^\sigma\mathbf{S}^{-1}
  + \mathbf{S}^{-1}\mathbf{K}^\sigma\mathbf{P}^\sigma \right).
\end{equation}
The final formula for the conventional exact-exchange energy
density via the resolution of the identity is
\begin{equation}
 e_\mathrm{x\sigma}^\mathrm{ex(conv)}(\bfr)
 = \frac{1}{2} \sum_{\mu\nu} Q_{\mu\nu}^\sigma
 \chi_\mu(\bfr)\chi_\nu^*(\bfr).  \label{eq:ex-RI}
\end{equation}
In practice, $e_\mathrm{x\sigma}^\mathrm{ex(conv)}$ 
is computed using the subroutines that evaluate
$n(\bfr)$ by passing
$\frac{1}{2}\mathbf{Q}^\sigma$ in place of $\mathbf{P}^\sigma$.

\subsection{Evaluation of the exact-exchange energy density
in the TPSS gauge}

The exact-exchange energy density in the TPSS gauge
is given by Eq.~(\ref{eq:ex-tpss}). The first term,
$e_\mathrm{x\sigma}^\mathrm{ex(conv)}(\bfr)$,
is computed by Eq.~(\ref{eq:ex-RI}) and the
gauge term is evaluated as follows.
Let us rewrite Eq.~(\ref{eq:spindelta}) as
\begin{equation}
 G_\sigma(\bfr) = a \left[ \nabla f_\sigma(\bfr)\cdot\nabla
  \tilde{\varepsilon}_\sigma(\bfr)
  + f_\sigma(\bfr)\nabla^2 \tilde{\varepsilon}_\sigma(\bfr) \right],
\end{equation}
where
\begin{equation}
 f_\sigma(\bfr) = \frac{n_\sigma/\tilde{\varepsilon}_\sigma^{2}}
 {1 + 4c\left(n_\sigma/\tilde{\varepsilon}_\sigma^3\right)^2}
 \left(\frac{\tau_\sigma^W}{\tau_\sigma}\right)^b.
\end{equation}
Based on Eq.~(\ref{eq:vareps}),
\begin{equation}
 \nabla \tilde{\varepsilon}_\sigma 
 = -\frac{\nabla e_\mathrm{x\sigma}^\mathrm{ex(conv)}
 + \tilde{\varepsilon}_\sigma \nabla n_\sigma}{n_\sigma},
  \label{eq:deps}
\end{equation}
\begin{eqnarray}
 \nabla^2 \tilde{\varepsilon}_\sigma
 & = & -\frac{\nabla^2 e_\mathrm{x\sigma}^\mathrm{ex(conv)}
 + 2\nabla\tilde{\varepsilon}_{\sigma}\cdot\nabla n_\sigma
 + \tilde{\varepsilon}_{\sigma}\nabla^2n_\sigma}{n_\sigma}.
  \label{eq:d2eps}
\end{eqnarray}
Eqs.~(\ref{eq:deps}) and~(\ref{eq:d2eps}) involve the first
and second derivatives of the exact-exchange
energy density in the conventional gauge.
These quantities are computed as
\begin{equation}
 \nabla e_\mathrm{x\sigma}^\mathrm{ex(conv)}
 = \frac{1}{2} \sum_{\mu\nu} Q_{\mu\nu}^\sigma
  \nabla[\chi_\mu(\bfr)\chi_\nu^*(\bfr)],  \label{eq:dex}
\end{equation}
\begin{equation}
 \nabla^2 e_\mathrm{x\sigma}^\mathrm{ex(conv)}
 = \frac{1}{2} \sum_{\mu\nu} Q_{\mu\nu}^\sigma
  \nabla^2[\chi_\mu(\bfr)\chi_\nu^*(\bfr)], \label{eq:d2ex}
\end{equation}
using the same subroutines that evaluate $\nabla
n_\sigma(\bfr)$ and $\nabla^2n_\sigma(\bfr)$ by passing
$\frac{1}{2}\mathbf{Q}^\sigma$ instead of 
$\mathbf{P}^\sigma$. Note that Eq.~(\ref{eq:ex-RI1}) clearly shows
that matrix elements $Q_{\mu\nu}^\sigma$ themselves do not depend
on $\bfr$.

\begin{figure}
\includegraphics[width=\columnwidth]{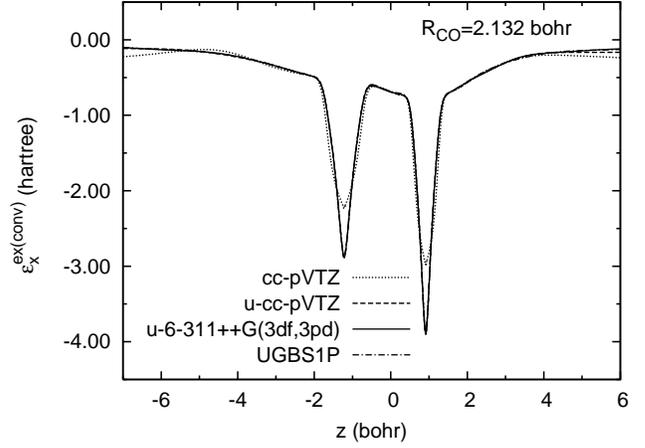}
\caption{\label{fig:co}
Conventional exact-exchange energy per electron in
the CO molecule along the internuclear axis evaluated at the
experimental geometry using the approximate resolution of
the identity in four basis sets. The curves for the last three
basis sets are close together everywhere.
}
\end{figure}

\begin{figure}
\includegraphics[width=\columnwidth]{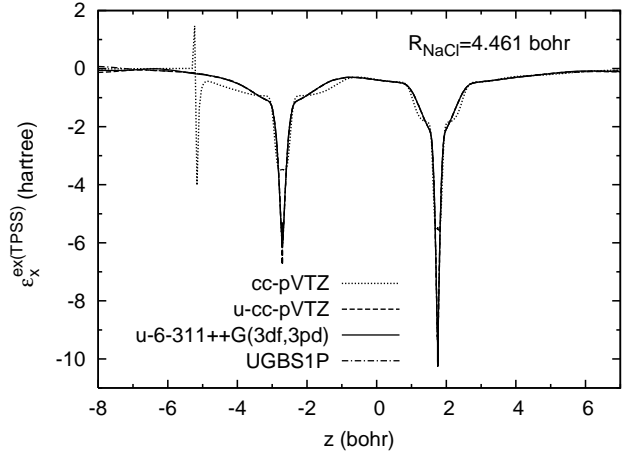}
\caption{\label{fig:nacl}
Exact-exchange energy per electron in the TPSS
gauge in the NaCl molecule evaluated at the experimental
geometry using the approximate resolution of the identity in
four basis sets. The curves for the last three basis sets are
almost indistinguishable.
}
\end{figure}

The gradient $\nabla f_\sigma(\bfr)$ can be written as
\begin{equation}
 \nabla f_\sigma(\bfr) = 
  \left(\frac{\tau_\sigma^W}{\tau_\sigma}\right)^b
 \nabla g_\sigma(\bfr) + g_\sigma(\bfr)
  \nabla \left(\frac{\tau_\sigma^W}{\tau_\sigma}\right)^b,
\end{equation}
where
\begin{equation}
 g_\sigma(\bfr) = \frac{n_\sigma \tilde{\varepsilon}_\sigma^{4}}
 {\tilde{\varepsilon}_\sigma^6 + 4cn_\sigma^2}.
\end{equation}
The quantity $\nabla g_\sigma(\bfr)$ is evaluated using the chain rule
as usual and it involves only the first derivatives of
$n(\bfr)$ and $e_\mathrm{x\sigma}^\mathrm{ex(conv)}(\bfr)$.
The second term can be written as
\begin{equation}
 \nabla \left(\frac{\tau_\sigma^W}{\tau_\sigma}\right)^b
 = b \left(\frac{\tau_\sigma^W}{\tau_\sigma}\right)^{b}
 \left( \frac{\nabla\tau_\sigma^W}{\tau_\sigma^W}
  - \frac{\nabla\tau_\sigma}{\tau_\sigma} \right).
\end{equation}
The gradient $\nabla\tau_\sigma^W$
involves derivatives of the type
\begin{equation}
 \frac{\partial|\nabla n_\sigma|^2}{\partial x}
 = 2\left( \frac{\partial n_\sigma}{\partial x}
  \frac{\partial^2 n_\sigma}{\partial x^2}
 + \frac{\partial n_\sigma}{\partial y}
  \frac{\partial^2 n_\sigma}{\partial y\partial x}
 + \frac{\partial n_\sigma}{\partial z}
  \frac{\partial^2 n_\sigma}{\partial z\partial x}
 \right)  \label{eq:dgam-dx}
\end{equation}
and similar expressions for $\partial|\nabla n_\sigma|^2/\partial y$ and
$\partial|\nabla n_\sigma|^2/\partial z$.
Finally, the gradient of the Kohn-Sham kinetic energy density
$\nabla\tau_\sigma$ has the components
\begin{equation}
 \frac{\partial\tau_\sigma}{\partial x}
 = \sum_i^\mathrm{occ.}
 \left( \frac{\partial\phi_{i\sigma}}{\partial x}
  \frac{\partial^2\phi_{i\sigma}}{\partial x^2}
 + \frac{\partial\phi_{i\sigma}}{\partial y}
  \frac{\partial^2\phi_{i\sigma}}{\partial y\partial x}
 + \frac{\partial\phi_{i\sigma}}{\partial z}
  \frac{\partial^2\phi_{i\sigma}}{\partial z\partial x}
 \right) \label{eq:dtau-dx}
\end{equation}
and similarly for $\partial\tau_\sigma/\partial y$
and $\partial\tau_\sigma/\partial z$.

The quantities given by Eqs.~(\ref{eq:dgam-dx})
and~(\ref{eq:dtau-dx}) are not used in any of the common GGA
and meta-GGA functionals and may not be immediately available in
standard density functional codes. However, the first and second
derivatives of the orbitals, from which Eqs.~(\ref{eq:dgam-dx})
and~(\ref{eq:dtau-dx}) are built, are readily available.  Thus,
evaluation of the exact-exchange energy density and the gauge
correction requires some modification of existing subroutines.
We have implemented these formulas in a development version
of the \textsc{gaussian} program~\cite{GDV-D1+}.

\subsection{Basis set effects}

We use the same nonorthogonal basis set $\{\chi_\mu\}$ to expand
the Kohn-Sham orbitals and to approximate the $\delta$-function by
Eq.~(\ref{eq:delta-f}).  Since Eq.~(\ref{eq:delta-f}) in a finite
basis set is not exact, the conventional exact-exchange energy
density $e_\mathrm{x}^\mathrm{ex(conv)}$ and its derivatives are only
approximate when evaluated by Eqs.~(\ref{eq:ex-RI}), (\ref{eq:dex}),
and~(\ref{eq:d2ex}). In fact, small and medium-size contracted basis
sets may cause large errors in $e_\mathrm{x}^\mathrm{ex(conv)}$
that are further magnified in $e_\mathrm{x}^\mathrm{ex(TPSS)}$
via $\nabla e_\mathrm{x}^\mathrm{ex(conv)}$ and $\nabla^2
e_\mathrm{x}^\mathrm{ex(conv)}$. Figs.~\ref{fig:co}
and~\ref{fig:nacl} show that, for instance, the cc-pVTZ
basis is insufficiently flexible. On the other hand, the uncontracted
cc-pVTZ basis set works almost as well as the near-complete UGBS1P
basis~\cite{Castro:1998/JCP/5225,G03:UserGuide}. In general,
uncontracted basis sets work much better in resolution of the
identity techniques than the corresponding contracted bases.

Furthermore, when cuspless Gaussian-type basis functions are
used, $|\nabla n_\sigma|^2$ and $|\nabla\tau_\sigma|^2$ exhibit
spurious oscillations in the vicinity of a nucleus.  However,
these artifacts are common to all semilocal density functional
calculations employing Gaussian-type orbitals, are negligible
energetically and may be ignored.

In summary, we caution against using medium-size contracted
basis sets like cc-pVTZ or 6-311+G* in Eqs.~(\ref{eq:ex-RI}),
(\ref{eq:dex}), and~(\ref{eq:d2ex}).  When in doubt, it is always
safer to uncontract the basis set.  In particular, we recommend
the fully uncontracted 6-311++G(3df,3pd) basis set, denoted as
u-6-311++G(3df,3pd), which strikes perfect balance between accuracy
and computational cost.

\section{Results}

\begin{figure}
\includegraphics[width=\columnwidth]{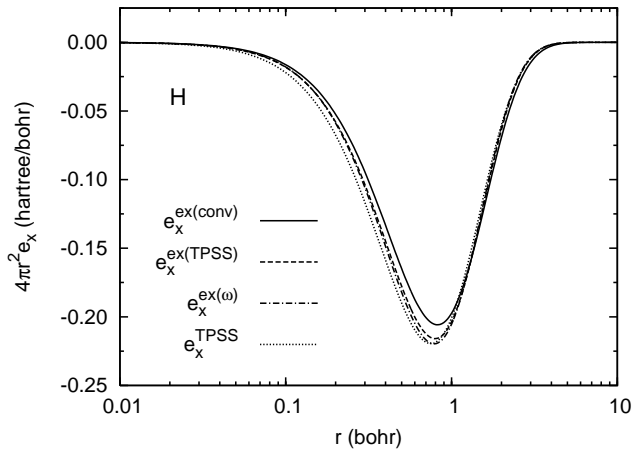}
\caption{\label{fig:h}
Radial exchange energy densities of the H atom computed
at the exact ground-state density:
exact conventional [ex(conv)], exact in the TPSS gauge [ex(TPSS)],
exact from a transformed exchange hole [ex($\omega$), $\omega=0.92$],
and TPSS.}
\end{figure}

\begin{figure}
\includegraphics[width=\columnwidth]{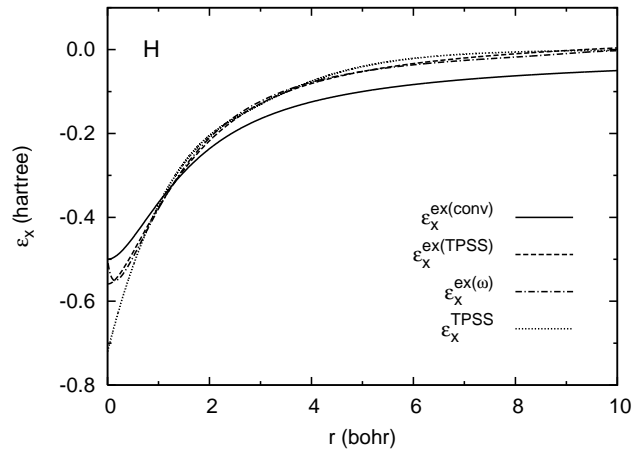}
\caption{\label{fig:h-eps}
Exchange energies per electron in the H atom computed
at the exact ground-state density. For the explanation
of the legend, refer to Fig.~\ref{fig:h}.
}
\end{figure}

\begin{figure}
\includegraphics[width=\columnwidth]{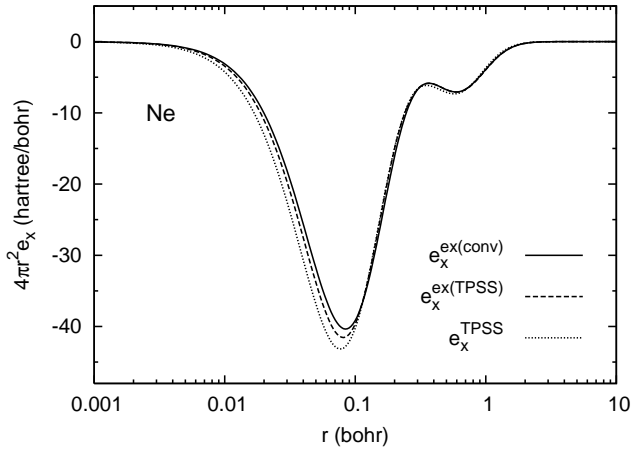}
\caption{\label{fig:ne}
Radial exchange energy densities of the Ne atom computed
at the converged Hartree-Fock orbitals in the UGBS basis set:
exact conventional [ex(conv)], exact in the TPSS gauge [ex(TPSS)],
and semilocal TPSS approximation.}
\end{figure}

\begin{figure}
\includegraphics[width=\columnwidth]{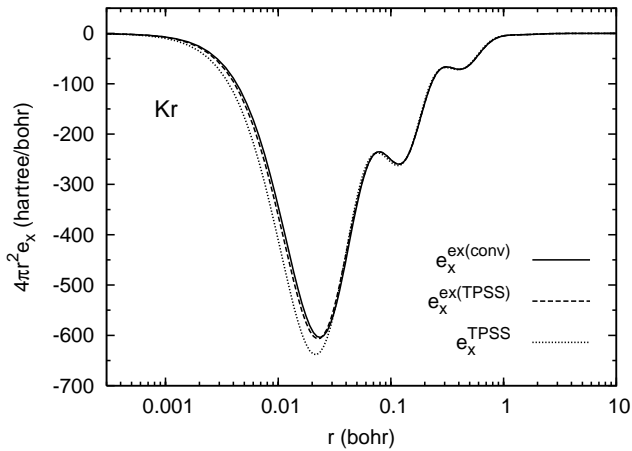}
\caption{\label{fig:kr}
Same as in Fig.~\ref{fig:ne} for the Kr atom.}
\end{figure}

\begin{figure}
\includegraphics[width=\columnwidth]{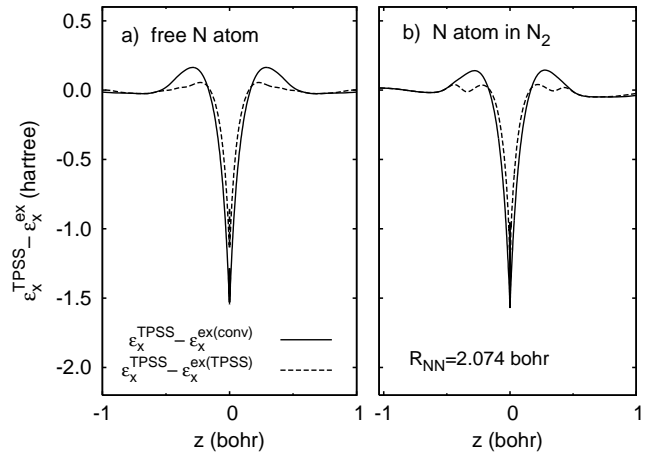}
\caption{\label{fig:nn2}
Difference $\varepsilon_\mathrm{x}^\mathrm{TPSS}
-\varepsilon_\mathrm{x}^\mathrm{ex}$, where
$\varepsilon_\mathrm{x}^\mathrm{ex}$
is in the conventional and TPSS gauges,
in a free N atom and along the internuclear axis of the N$_2$ molecule
at the experimental geometry. Panel b) shows only the right half
of the molecule with the N nucleus placed at $z=0$.
All quantities were computed at
the converged TPSS orbitals using the UGBS1P basis set.}
\end{figure}

\begin{figure}
\includegraphics[width=\columnwidth]{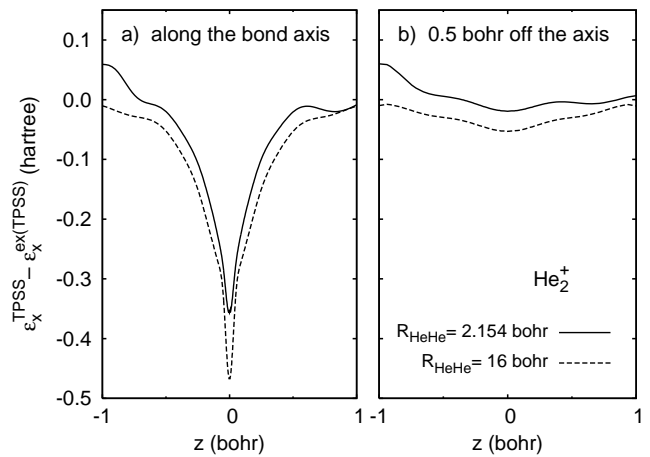}
\caption{\label{fig:he2+}
Difference $\varepsilon_\mathrm{x}^\mathrm{TPSS}
-\varepsilon_\mathrm{x}^\mathrm{ex(TPSS)}$ in the He$_2^{+}$ molecule
along the internuclear axis and along a parallel axis offset by
0.5 bohr. Each panel shows only the region near the
right nucleus which is always placed at $z=0$.  The static correlation in
the stretched molecule He$^{0.5+}\cdots$He$^{0.5+}$ (dashed
line) is more negative than at the equilibrium TPSS/cc-pVQZ geometry
(solid line). All quantities were computed at the converged TPSS
orbitals using the uncontracted cc-pVQZ basis set.}
\end{figure}

The fact that the TPSS meta-GGA was designed to recover many
exact properties~\cite{Tao:2003/PRL/146401,Staroverov:2004/PRB/075102}
of the exact-exchange functional does
not guarantee that $e_\mathrm{x}^\mathrm{TPSS}$ is close to
$e_\mathrm{x}^\mathrm{ex(conv)}$. This is evident from Fig.~\ref{fig:h}
which shows radial plots of these energy densities
in the H atom. The TPSS and conventional exact-exchange energy
density are different, even though they both integrate to the same
exact value of $-5/16$ hartree~\cite{Tao:2003/PRL/146401}.
The exact-exchange energy densities in the TPSS gauge
$e_\mathrm{x}^\mathrm{ex(TPSS)}$ and $e_\mathrm{x}^\mathrm{ex(\omega)}$
are both much closer than $e_\mathrm{x}^\mathrm{ex(conv)}$ to
the semilocal $e_\mathrm{x}^\mathrm{TPSS}$.

Fig.~\ref{fig:h-eps} shows the exchange energy per electron,
$\varepsilon_\mathrm{x}=e_\mathrm{x}/n$ vs.~$n$ for the H atom,
comparing the exact conventional, exact in the TPSS gauge,
and the hole-transformed ($\omega=0.92$) exact-exchange
energies per electron to the semilocal TPSS exchange
approximation. Unlike Fig.~\ref{fig:h}, this figure shows what
happens in the energetically unimportant small-$r$ and large-$r$
regions. The transformed exact-exchange energy per electron
$\varepsilon_\mathrm{x}^\mathrm{ex(\omega)}$ stands apart from the
others in that it appears to have an inverted cusp at the nucleus.

Figs.~\ref{fig:ne} and~\ref{fig:kr} compare
exchange energy in the Ne and Kr atoms
evaluated in a post-self-consistent manner at the
converged Hartree-Fock orbitals obtained using the near-complete
UGBS basis set~\cite{Castro:1998/JCP/5225}. Overall, 
Figs.~\ref{fig:h}--\ref{fig:kr} suggest that, for the smaller atoms,
the exact-exchange energy density in the TPSS gauge is closer
than $e_\mathrm{x}^\mathrm{ex(conv)}$ to the TPSS exchange energy density.
For larger atoms, however, $e_\mathrm{x}^\mathrm{ex(TPSS)}$ remains
closer to $e_\mathrm{x}^\mathrm{ex(conv)}$ than to 
$e_\mathrm{x}^\mathrm{TPSS}$ in the deep core region.

Fig.~\ref{fig:nn2} shows that the exact-exchange energy density
(per electron) in the gauge of a semilocal approximation,
$\varepsilon_\mathrm{x}^\mathrm{ex(TPSS)}$, differs
from $\varepsilon_\mathrm{x}^\mathrm{ex(conv)}$
in a non-trivial way. The difference
$\varepsilon_\mathrm{x}^\mathrm{TPSS}-\varepsilon_\mathrm{x}^\mathrm{ex(TPSS)}$
reveals subtle effects in the N$_2$ molecule
that are absent in a free N atom.  Off-axis effects (not
shown) are of course important.  In contrast, the difference
$\varepsilon_\mathrm{x}^\mathrm{TPSS}-\varepsilon_\mathrm{x}^\mathrm{ex(conv)}$
in the N$_2$ molecule is very similar to that in a free N atom.

Fig.~\ref{fig:he2+} shows that the difference
$\varepsilon_\mathrm{x}^\mathrm{TPSS}-\varepsilon_\mathrm{x}^\mathrm{ex(TPSS)}$
representing the static correlation gets substantially more negative
upon bond stretching, when the fragments show large fluctuations
of electron number at the Hartree-Fock level, as it should.
Note that the nuclei in the stretched He$_2^{+}$ molecule
(R$_\mathrm{HeHe}=16$ bohr) are essentially isolated, so the difference
$\varepsilon_\mathrm{x}^\mathrm{TPSS}-\varepsilon_\mathrm{x}^\mathrm{ex(TPSS)}$
is almost perfectly symmetric about $z=0$.

\begin{figure}
\includegraphics[width=\columnwidth]{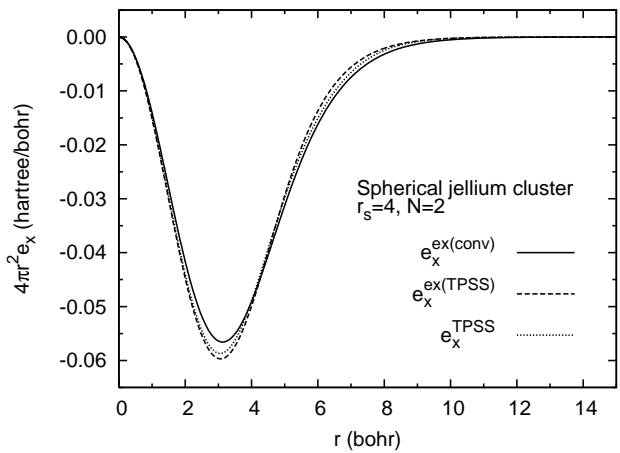}
\caption{\label{fig:j2}
Radial exchange energy densities for a spherical jellium
cluster of $N=2$ electrons computed at the exchange-only
OEP orbitals: exact conventional [ex(conv)], exact in the TPSS gauge
[ex(TPSS)], and semilocal (TPSS).}
\end{figure}

\begin{figure}
\includegraphics[width=\columnwidth]{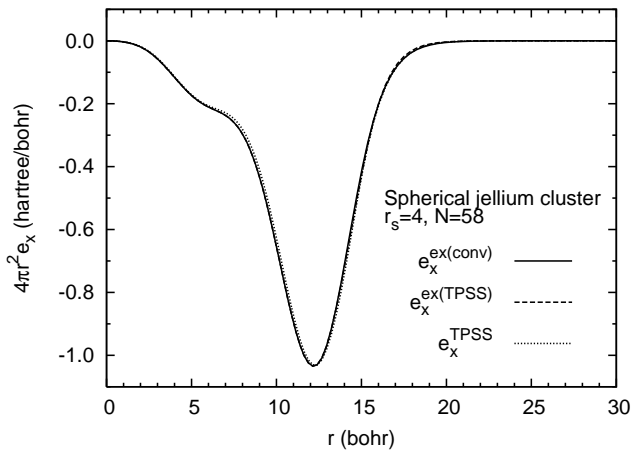}
\caption{\label{fig:j58}
Same as in Fig.~\ref{fig:j2} for a cluster of
$N=58$ electrons.}
\end{figure}

We have also evaluated the exact-exchange energy density in
the TPSS gauge and the two conventional energy densities for
spherical jellium clusters. A spherical jellium cluster is
a model system that has a uniform positive background charge
and a spherically distributed electron density. The radius of
the sphere is given by $R = r_s N^{1/3}$, where $r_s$ is the
bulk density parameter and $N$ is the number of electrons in
the system. The volume of sphere is proportional to $N$ and is
given by the relation $V = (4\pi/3)Nr_s^3$. The three exchange
energy densities for the jellium clusters of $r_s = 4$ for $N=2$
and 58 are shown in Figs.~\ref{fig:j2} and~\ref{fig:j58}. All
three quantities, $e_\mathrm{x}^\mathrm{ex(conv)}$,
$e_\mathrm{x}^\mathrm{ex(TPSS)}$, and $e_\mathrm{x}^\mathrm{TPSS}$,
were evaluated at the orbitals and densities obtained from OEP
calculations~\cite{Talman:1976/PRA/36,Engel:1999/JCC/31,Kummel:2003/PRL/043004}.
As with atoms and molecules, $e_\mathrm{x}^\mathrm{ex(TPSS)}$
is closer than $e_\mathrm{x}^\mathrm{ex(conv)}$
to $e_\mathrm{x}^\mathrm{TPSS}$ in these jellium clusters.

\section{Conclusion}

As observed before~\cite{Folland:1971/PRA/1535,Becke:1988/PRA/3098,%
Tao:2001/JCP/3519}, the exchange energy density of a semilocal
functional is reasonably close to the conventional exact-exchange
energy density of Eq.~(\ref{eq:ex-conv}) in compact systems like
atoms or spherical jellium clusters.  We confirm this here for the
nonempirical TPSS meta- GGA. The relative differences are largest in
regions of space where the density is dominated by a single orbital
shape, making $\tau^W/\tau$ close to 1, e.g., the H or He atoms and
the two-electron jellium cluster. Particularly in these regions,
the difference can be reduced by a gauge trans- formation of the
conventional exact-exchange energy den- sity.

We have found a simple, realistic, and not too highly
parametrized form of the function $G(\bfr)$, given by
Eq.~(\ref{eq:delta}), which via Eq.~(\ref{eq:ex-tpss}) transforms
the conventional difference of semilocal and exact-exchange energy
densities appearing in Eq.~(\ref{eq:lh}) to the gauge of
the TPSS meta-GGA. This transformation solves the problem of
nonuniqueness of the exact-exchange energy density arising in the
context of modeling the static correlation by the difference of
semilocal and exact exchange energy densities.  The transformed
exact-exchange energy density $e_\mathrm{x}^\mathrm{ex(TPSS)}$
does in fact contain more information about electron correlation
than $e_\mathrm{x}^\mathrm{ex(conv)}$. In a forthcoming
article~\cite{HGGA:xxx}, we will present a construction of a
hyper-GGA that relies on this gauge transformation to give highly
accurate thermochemistry and reaction barriers.

Finally, we have demonstrated that, as
expected~\cite{Perdew:2007/PRA/040501}, the difference between
semilocal and exact-exchange energy densities becomes more negative
under bond stretching in He$_2^{+}$ and related systems, where the
separating fragments show large fluctuations of electron number at
the independent-electron level.

\begin{acknowledgments}
This work was supported by the NSF under Grants DMR-0501588
(J.T. and J.P.P.) and CHE-0457030 (V.N.S. and G.E.S.), by
DOE under Contract No. DE-AC52-06NA25396 and the LDRD programs at LANL (J.T.),
and by the NSERC of Canada (V.N.S.)
\end{acknowledgments}

\end{document}